\documentclass[aps,prl,showpacs,twocolumn,notitlepage]{revtex4-1}
\usepackage{amsmath,bm,epsfig,graphics,txfonts,subeqnarray,setspace,graphicx,epstopdf,subfigure,color}

\def\b{\mathbf}

\def\Xint#1{\mathchoice
{\XXint\displaystyle\textstyle{#1}}%
{\XXint\textstyle\scriptstyle{#1}}%
{\XXint\scriptstyle\scriptscriptstyle{#1}}%
{\XXint\scriptscriptstyle\scriptscriptstyle{#1}}%
\!\int}
\def\XXint#1#2#3{{\setbox0=\hbox{$#1{#2#3}{\int}$}
\vcenter{\hbox{$#2#3$}}\kern-.5\wd0}}

\def\dashint{\Xint-}

\def\H{\mathcal{H}}
\def\V0{V_0}
\def\kap{\kappa}
\def\tkap{\tilde{\kappa}}

\begin{document}
\title{Active matter invasion of a viscous fluid: unstable sheets and a no-flow theorem}
\author{Christopher J. Miles}
\email{cmiless@umich.edu}
\affiliation{Department of Physics, University of Michigan, 450 Church St., Ann Arbor, MI 48109, USA}
\author{Arthur A. Evans}
\affiliation{Department of Mathematics, University of Wisconsin-Madison, 480 Lincoln Dr., Madison, WI 53706, USA}
\author{Michael J. Shelley}
\affiliation{Flatiron Institute, Simons Foundation, New York, NY, USA; and Courant Institute of Mathematical Sciences, New York University, New York, NY 10012, USA}
\author{Saverio E. Spagnolie}
\email{spagnolie@math.wisc.edu}
\affiliation{Department of Mathematics, University of Wisconsin-Madison, 480 Lincoln Dr., Madison, WI 53706, USA }

\date{\today}

\begin{abstract}
We investigate the dynamics of a dilute suspension of hydrodynamically interacting motile or immotile stress-generating swimmers or particles as they invade a surrounding viscous fluid. Colonies of aligned pusher particles are shown to elongate in the direction of particle orientation and undergo a cascade of transverse concentration instabilities, governed at small times by an equation which also describes the Saffman-Taylor instability in a Hele-Shaw cell, or Rayleigh-Taylor instability in two-dimensional flow through a porous medium. Thin sheets of aligned pusher particles are always unstable, while sheets of aligned puller particles can either be stable (immotile particles), or unstable (motile particles) with a growth rate which is non-monotonic in the force dipole strength. We also prove a surprising ``no-flow theorem'': a distribution initially isotropic in orientation loses isotropy immediately but in such a way that results in no fluid flow {\it everywhere and for all time.}
\end{abstract}

\maketitle

The last decade has seen an explosion of interest in the collective dynamics of active particles immersed in fluids, from swimming microorganisms to magnetically driven and phoretic colloidal particles \cite{pk92,dccgk04,ccdgk07,uhg08,bm09,Ramaswamy10,ks11,vz12,mjrlprs13,ss15,ewg15,zs16,ylp16} to kinesin-driven microtubule assemblies \cite{hhw13,scdhd12,klsdgbmdb14,pjj15,ffsn15,gbgbs15,Shelley16,mmgm18}. A first-principles model of active suspensions is a mean-field kinetic theory which tracks the distribution of particle positions and orientations and which may include hydrodynamic interactions \cite{ss07,ss08,sk09,ss15,ks11} and short-range physics \cite{mn08,sk09}. Constituent particles are classified as either ``pushers'' or ``pullers'' depending on the sign of the generated stresslet flow, which in turn depends on the geometry of the body and the mechanism of stress-generation \cite{ss08,lp09,sl12,ga14,lm16,ne18}. Other models range from Landau-de Gennes ``Q tensor'' theories to moment-closure theories \cite{rst03,wg12,fwz13,gbmm13,tgy13,bclb13}. Generic features in these systems include long-range coherence, topological defects, and instability \cite{sr02,ss08,hs10,bcgmpr13,ess13,bclb13,scm14}.

Much is known about active suspensions which cover the entire available physical domain. Far less is known about the invasion of a surrounding particle-free environment, though this is of considerable importance in the dynamic self assembly of swarms \cite{sd97,cw09,dtrb10}, and in the formation of biofilms, mycelia, and fruiting bodies \cite{crksv14}. Novel means of bringing bacteria into a confined region using external flows have allowed for a closer look at rapid expansion, including acoustic-trapping \cite{tdvb16,ghr18}, UV-light exposure \cite{pga18} and vortical flows \cite{sa16,srba18}. The effects of confinement by soft boundaries with surface tension has seen theoretical treatment \cite{gl17,gbjs17}, and unstable bands of active particles have been studied in a dry system \cite{npabgc14}.

In this Letter we investigate the dynamics of colonies (a coherent collective) of either motile or immotile active particles as they invade a surrounding viscous fluid. Colonies of aligned pushers are shown to elongate in the direction of particle orientation and then undergo a cascade of transverse concentration instabilities. The initial instability in two-dimensions is shown to be governed at small times by an equation which also describes the Saffman-Taylor instability in a Hele-Shaw cell (flow through a small gap between two nearby plates), or the Rayleigh-Taylor instability in two-dimensional flow through a porous medium, respectively. Linear stability analysis offers approximations that match the results of full numerical simulations. We close with a proof and demonstration of a counter-intuitive ``no-flow theorem,'' that an isotropically oriented distribution with any initial concentration profile results in no fluid flow everywhere and for all time. 

\noindent \textbf{Mathematical model:} Following Refs.~\cite{ss08,sk09}, we describe a dilute suspension of $N$ self-propelled rod-like particles in a viscous fluid by the particle distribution function, $\Psi(\b{x},\b{p},t)$, where $\b{x}$ is the particle position in a periodic spatial domain $D$ while $\b{p}$ is the particle orientation vector on the unit ball $S$ ($|\b{p}|=1$). The number of particles is conserved, $N=\int_{D}\int_{S} \Psi( \b{x},\b{p},t)  d\b{p} d\b{x}$, resulting in a Smoluchowski equation, 
\begin{gather}
\label{eq:smoluchowski}
\Psi_t + \nabla\cdot \left(\dot{\b{x}}\Psi\right)+ \nabla_{\b{p}}\cdot \left(\dot{\b{p}}\Psi\right) = 0,
\end{gather}
where $\nabla=\nabla_\b{x}$ and $\nabla_{\b{p}} = (\b{I}-\b{p}\b{p})\cdot \partial/\partial \b{p}$. Neglecting collisions~\cite{ess13}, the fluxes $\dot{\b{x}}$ and $\dot{\b{p}}$ are given by
\begin{gather}
\dot{\b{x}}=V_{0}\b{p}+\b{u}(\b{x})-d_t \nabla \left(\ln \Psi\right),\\
\dot{\b{p}}=(\b{I-pp})\cdot (\b{p}\cdot \nabla \b{u})-d_r \nabla_{\b{p}}\left(\ln \Psi\right),
\end{gather}
with $V_0$ the swimming speed, $d_{t}$ ($d_{r}$) the translational (rotational) diffusivity, $\b{u}(\b{x},t)$ the fluid velocity, and $\b{pp}$ a dyadic product.

 \begin{figure*}[t]
\begin{center}
\includegraphics[width=\textwidth]{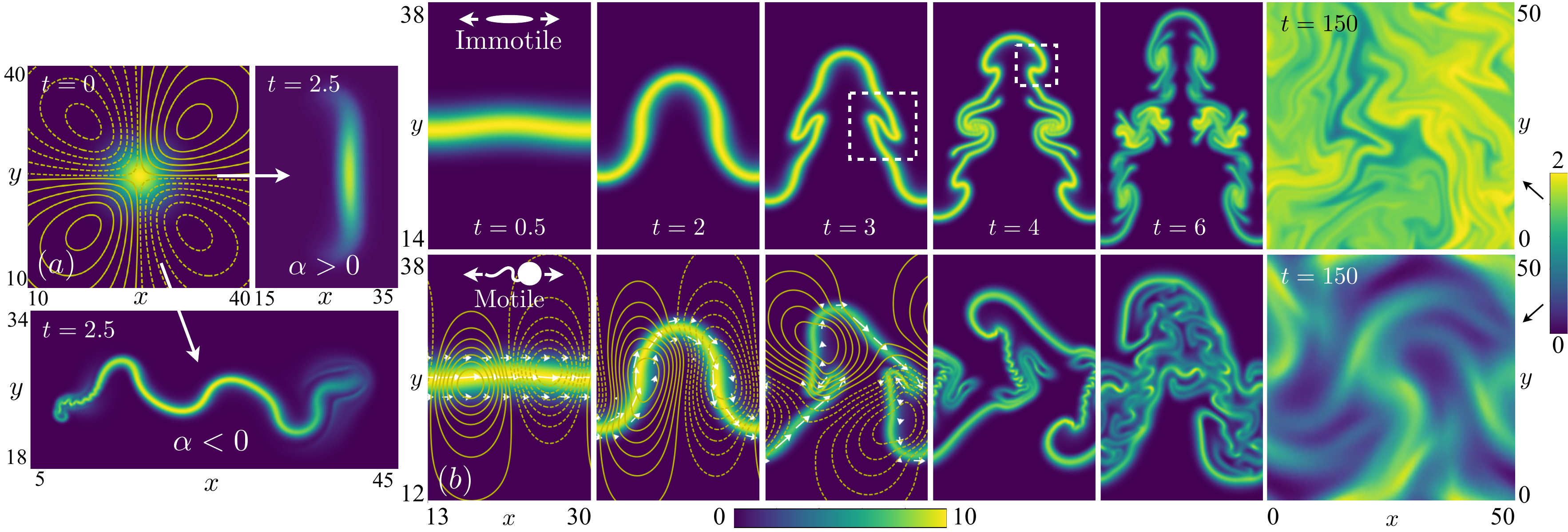}
\vspace{-.1in}
\caption{(a) Concentration evolution of a weakly perturbed cylindrical colony (Gaussian in cross-section) of aligned motile pullers ($\alpha=0.5$, right) and pushers ($\alpha=-0.5$, bottom). Streamfunction contours are included at $t=0$ with solid (dashed) lines representing positive (negative) values if $\alpha<0$, and signs reversed if $\alpha>0$. See Supplementary Movie M1. (b) Evolution of a thin band of immotile (top) and motile (bottom) pusher particles with $\alpha=-1$, with initial distribution function $\Psi(\b{x},\theta,t=0)=C\exp\{-(y-h(x))^2/a^2-\theta^2/b^2\}$ and $C$ a normalization constant. The initial perturbation is given by $h(x)=0.1\sin(6\pi x/L)$ and $(a,b)=(2,0.2)$. The polarity field $\langle \b{p} \rangle$ (arrows) shows the local particle orientation. Exponential growth in amplitude leads to a secondary instability and self-folding at $t\approx 2.5$, which plays out again at $t\approx 4$ with similar features on a smaller scale (dashed boxes). The small initial spread in orientation and small noise results in eventual loss of symmetry and the system arrives in an unsteady roiling state $(t=150)$ with uniform concentration (immotile particles) or concentration bands (motile particles). See Supplementary Movie M2.}
\label{fig:MainFig}
\end{center}
\vspace{-.25in}
\end{figure*}

The environment is assumed to be a viscous Newtonian fluid, driven by stresses generated by the suspended particles. The flow field $\b{u}$ satisfies the Stokes equations, consisting of momentum and mass conservation, 
\begin{gather}\label{eq:Stokes}
-\nabla q + \mu \nabla^2 \b{u} + \nabla \cdot \bm{\Sigma}_a=\b{0},\,\,\,\, \nabla \cdot \b{u}=0,
\end{gather}
with $q$ the pressure, $\mu$ the dynamic viscosity, and $\bm{\Sigma}_a=\sigma\langle \b{pp} \rangle$ the active stress (proportional to the second orientational moment of $\Psi$, see below). The coefficient $\sigma$ is the force dipole (or stresslet) strength, with $\sigma<0$ for pushers and $\sigma>0$ for pullers \cite{ss15}, which has been computed for ellipsoidal ``Janus'' particles~\cite{sl12,lm16} and for more general particle types \cite{pl15,rkbh18,ne18}, and has been obtained experimentally for a few types of swimming cells \cite{dgmpt10,ddcgg11}. Orientational moments will be denoted by $\langle h(\b{p}) \rangle=\int_{S} h(\b{p})\Psi\,d\b{p}$. For example, integrating Eq.~\eqref{eq:smoluchowski} gives an evolution equation for the particle concentration, $c=\langle 1 \rangle$, of the form
\begin{gather}\label{eq:c}
c_t +\nabla \cdot( c\b{u}) - d_t \nabla^2 c =-V_0 \nabla \cdot \langle \b{p} \rangle,
\end{gather}
where $\langle \b{p}\rangle$ is the polarity \cite{ss15}.

With $\ell$ the active particle length, we scale velocities by the swimming speed, $V_0$, and lengths by the mean free path $\ell_{c}=(V/V_p)\ell$, where $V$ is the total fluid volume and $V_p=N \ell^3$ is an effective volume of particles, hence $\ell_c = V(N\ell^2)^{-1}$. Time is scaled by $\ell_c/V_0$, force densities are scaled by $\mu V_0/\ell_c^2$, and $\Psi$ is normalized by the particle number density, $N/V = (\ell^2\ell_c)^{-1}=(\ell/\ell_c)\ell^{-3}$. The dimensionless dipole strength is defined as $\alpha = \sigma/[\mu V_0 \ell_c^2]$. With all variables now taken to be dimensionless, particle conservation is written as $\tkap^{-1} \int_{D} \int_{S}\Psi \,d\b{p} d\b{x} =1$, where $\tkap^{-1}  = \ell_c^3/V$ is proportional to the particle volume fraction. In the case of immotile particles a different velocity scale is required \cite{SuppMat}. 

The far-field velocity due to an individual swimmer at the origin, oriented in the direction $\b{p}$, is $\b{u}=\alpha(8\pi)^{-1} \b{p}\b{p}:\nabla \b{G}(\b{x})$ where $G_{ij}=\delta_{ij}/|\b{x}|+x_i x_j/|\b{x}|^3$ is the Stokeslet singularity \cite{Pozrikidis92}. The active force density is then given by $\b{f}_a =\alpha(\ell_c/\ell)^2\nabla \cdot \langle \b{pp} \rangle$. Following Ref.~\cite{ss08} for the sake of comparison, we set $\ell_c/\ell=1$. The swimming speed, $\V0$, now taken as dimensionless, is unity for motile systems and zero otherwise.% and $\kap = L^3=50^3$}. %System dependrence on the swimming speed now appears only in the dimensionless dipole strength $\alpha$.

We will consider the case of confinement to motion in two-dimensions in a periodic domain $(x,y)\in [0,L)\times [0,H)$, and invariance in the $\hat{\b{z}}$ direction, writing $\b{p}=(\cos\theta,\sin\theta,0)$ and $\Psi(\b{x},\b{p},t)=\Psi(x,y,\theta,t)$. It is expedient to then define $\kap=HL$ so that $\kap^{-1} \int_0^L\int_0^H \int_{0}^{2\pi}\Psi \,d\theta\,dy\,dx =1$. Numerical solution of Eqs.~\eqref{eq:smoluchowski}-\eqref{eq:Stokes} using $L=H=50$ and a pseudospectral method with $256^3$ gridpoints and dealiasing (3/2 rule) \cite{Fornberg98}; an integrating factor method along with a second-order accurate Adams-Bashforth scheme is used for time-stepping.

\noindent \textbf{Dynamics of thin active sheets}:
To motivate the study to come we first consider the invasion of a concentrated cylindrical colony, Gaussian in cross-section, of motile particles initially aligned in the $\b{\hat{x}}$ direction into an empty viscous fluid, shown in Fig.~\ref{fig:MainFig}a. The associated global flow field is exactly that of a single regularized force dipole, resulting in the case of pullers in a stable concentration elongation in a direction orthogonal to the original swimmer orientation \cite{SuppMat}. For pushers the colony-induced velocity field changes sign and elongation is parallel to the swimming direction, but if slightly perturbed a transverse concentration instability ensues. Fore-aft symmetry is broken due to particle motility; the colony splays at the leading tip on the right, while undergoing a periodic folding at the rear reminiscent of the buckling of planar viscous jets \cite{cm81} and extruded beams \cite{gnp14} into viscous fluids.

To better understand this dramatic evolution we turn to the behavior of an infinite sheet of particles which are initially in alignment. Figure~\ref{fig:MainFig}b shows the evolution of a distribution of immotile (top) and motile (bottom) pusher particles, initially confined to a thin band and with a small transverse concentration perturbation. Early stages show rapid growth of the wave amplitude. At the same time individual particles are rotated towards the principal direction of the velocity gradient, so that they remain nearly tangent to the concentration band which results in a secondary instability and self-folding. The same structures are observed again and again on smaller length scales, though particle motility breaks left-right symmetry and significantly alters the structure of subsequent folding events. At longer times the system is finally drawn to an unsteady roiling state, with uniform concentration for immotile particles ($c$ satisfies a pure advection-diffusion equation in Eq.~\eqref{eq:c} in this case) or concentration bands described by Saintillan \& Shelley \cite{ss08} for motile particles. 

%Throughout the entire time simulated (after rapid initial relaxation) the Fourier spectrum of the concentration is extremely well approximated as exponentially decaying. Writing $c(\b{x},t)=\langle 1 \rangle = \sum_{\b{k}} \hat{c}_\b{k}(t)\exp(2\pi i \b{k}\cdot \b{x}/\sqrt{\kappa})$ we find $\hat{c}_\b{k} \approx \exp(-\gamma(t)|\b{k}|)$, with $\gamma(t)$ shown in Fig.~\ref{fig:MainFig}c. The highest available modes are excited at $t\approx 3$ when the secondary instability begins to fold the line density of particles onto itself. The decay rate recovers slowly for immotile particles, appearing to tend towards $\gamma \approx 0.150\pm 0.002$, and more rapidly for motile particles to $\gamma \approx 0.20 \pm 0.02$.}

%, with features similar to a slender-body theory of filaments in Stokes flow \cite{Lighthill76,Johnson80,lmss13,esbl13,la15,weg17}

To analyze the instability, let $h(x,t)$ and $\phi(x,t)$ represent the vertical displacement and polarity of the line distribution, respectively, with $\b{n}=\langle \b{p}\rangle/c = (\cos\phi,\sin\phi,0)$ the normalized polarity. We study the dynamics of this line distribution through its far-field self-influence. For small $h$ and $\phi$, and solving \eqref{eq:Stokes} for the velocity field \cite{SuppMat} we find 
\begin{gather}\label{eq:h}
h_t+\V0 h_x= v+\V0 \phi,\,\,\,\, \,\, \phi_t+\V0  \phi_x= v_x,\\
v=-\frac{\alpha\kappa}{4 L} \H[h_x],
%h_t+\V0 h_x= v^{(1)}(x,0,t)+\V0 \phi,\,\,\,\, \,\, \phi_t+\V0  \phi_x= v^{(1)}_x(x,0,t),
\end{gather}
where $v$ is the vertical component of velocity evaluated on the flat surface $h(x,t)=0$, and $\H[ \cdot ]$ is the Hilbert transform, 
\begin{gather}
\H[f](x)=\frac{1}{\pi}\dashint_{-\infty}^{\infty} \frac{f(y)}{x-y}\,dy.
\end{gather}

%To close the system we require the self-generated flow. Inserting a line distribution of particles into the active stress integral and carrying this through to the velocity field, for $\e \ll 1$ we find $v^{(1)}=-\ta \H[h_x]$, where $\ta = \alpha \kappa/(4\pi L)$ and $\H[ \cdot ]$ is the Hilbert transform, $\H[f]= \dashint_{0}^{L} f(x')/(x-x')\,dx'$ \cite{SuppMat}. 

%\begin{gather}\label{eq:PVint}
%\mathbf{u}(\b{x})=\frac{-\e \alpha \kap}{4\pi L} \dashint_{0}^{L} \frac{h(x,t)-h(x',t)}{(x-x')^2}\b{\hat{y}}\,dx'+O(\e^2),
%\end{gather}
%to be interpreted in the principal value sense \cite{SuppMat}. 

\begin{figure}[th]
\begin{center}
\includegraphics[width=.48\textwidth]{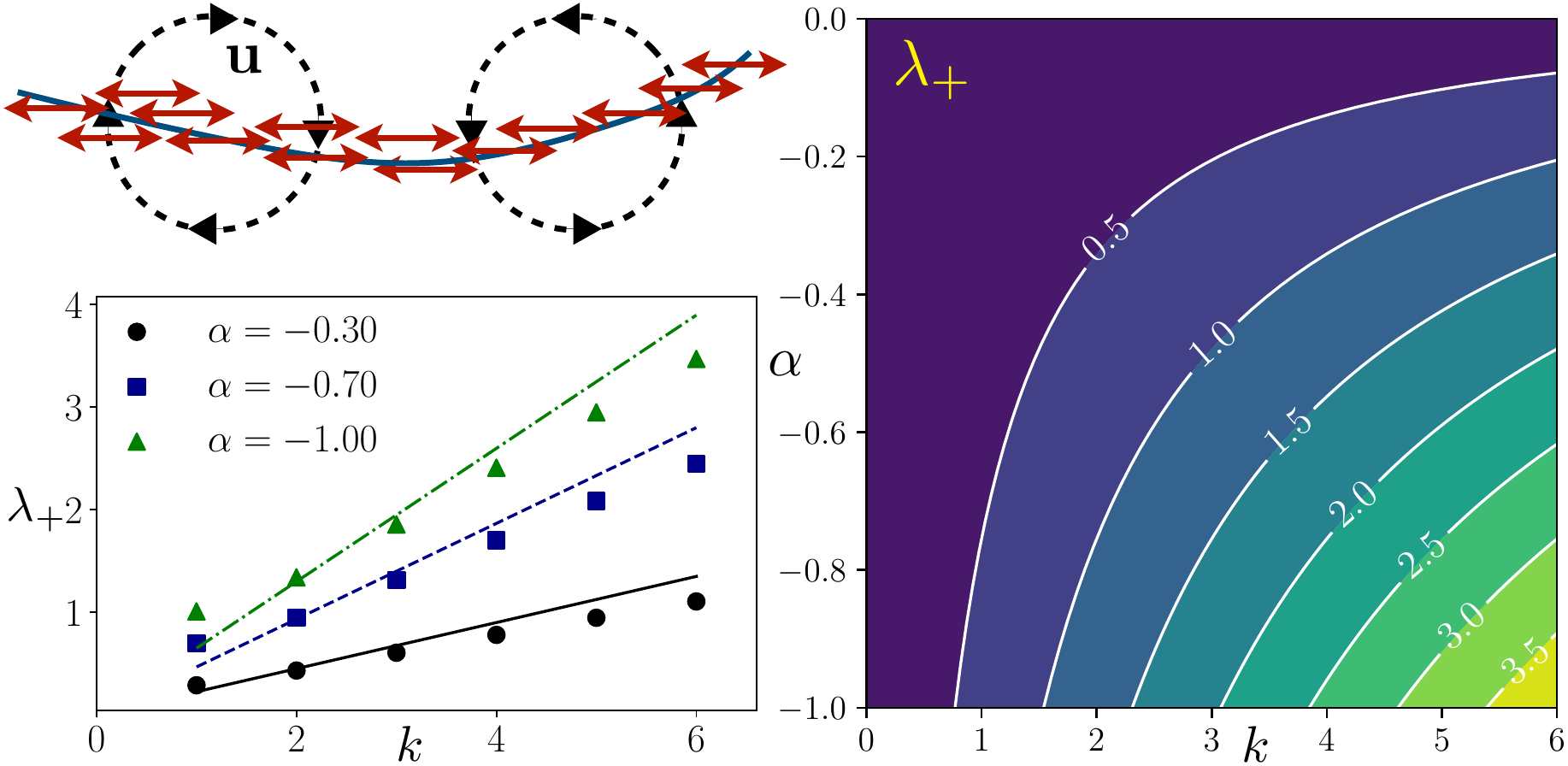}
\end{center}
\vspace{-.2in}
\caption{(Top left) A thin sheet of aligned pushers is unstable to transverse concentration perturbations due to its self-generated velocity field. (Bottom left) The positive growth rate for motile pushers, comparing numerics and theory. (Right) Theoretical growth rates for motile pushers for a range of dipole strengths and wave numbers.}
\label{fig:growth_rate}
\end{figure}

The Hilbert transform is diagonalized by changing to a Fourier basis, with $\H[e^{iqx}]=-i\, \mbox{sign}(q)e^{iqx}$. The Ansatz $(h(x,t),\phi(x,t))=\sum_k (\hat{h}_k(t),\hat{\phi}_k(t)) \exp(2\pi i k x/L)$ therefore results in a quadratic eigenvalue problem, and the eigenvalues
%\begin{gather}
%\b{u}(x,t) =\frac{-\pi  \e \alpha \kap}{2 L^2} \sum_k |k| \hat{h}_k(t) \exp(2\pi i k x/L) \b{\hat{y}}.
%\end{gather}
%Expressing $\phi$ similarly, the dynamics are governed at first order in $\e$ by
%\begin{gather}
%\hat{h}_k' +\frac{2\pi i k \V0}{L}\hat{h}_k= \frac{-\pi \alpha \kap   |k|}{2\mu L^2}\hat{h}_k+\V0 \hat{\phi}_k,\\
%\hat{\phi}_k' +\frac{2\pi i k \V0}{L}\hat{\phi}_k= \frac{-\pi^2 i k |k| \alpha \kap }{\mu L^3}  \hat{h}_k,
%\end{gather}
%a linear system with eigenvalues
\begin{gather}\label{eq:lambda}
\lambda_{\pm} = \frac{\pi}{4 L^4} \left(-\alpha \kap \left| k\right| -8 i k L \V0 \pm \gamma_k\right),
\end{gather}
where $\gamma_k = \sqrt{\alpha \kap  (\alpha \kap k^2-16 i L \V0 k |k|)}$. A comparison to numerics is shown in Fig.~\ref{fig:growth_rate} for motile pushers with three negative dipole strengths along with the theory for a wider range of $\alpha$ and $k$. The analytical predictions are accurate for the entire range studied, with discrepancies owing to the vertical periodic boundary condition and the non-vanishing thickness of the line distribution in the simulations.

In the immotile case, $V_0=0$ (or in the limit as $|\alpha|/\V0 \to \infty$), sheets of pushers are all unstable and sheets of pullers are all stable, with respective growth and decay rates both given by $-\pi \alpha \kappa |k|/(2 L^2)$. This behavior owes to the velocity field created by the active stress, illustrated in Fig.~\ref{fig:growth_rate} (see Supplementary Movie M2), which either amplifies or damps the initial concentration perturbation. The linearized dynamics are now governed solely by the equation 
\begin{gather}
h_t = \frac{-\alpha \kappa}{4L} \H[h_x].
\end{gather}
This expression establishes an unexpected connection to well-studied phenomena in entirely different settings: interfacial instabilities in gravity or pressure-driven Hele-Shaw problems, or two-dimensional flows in porous media, without surface tension, whose flow is governed by Darcy's Law \cite{ta83,hls94} (also known as the Muskat problem \cite{sch04,ccg11}). There, as in the present setting, the classical Saffman-Taylor or Rayleigh-Taylor instabilities are modified to exponential growth rate dependence which is linear in $|k|$ \cite{st58}. Such interfacial instabilities are associated with the formation of singularities in free-surface flows, for example the finite-time ``Moore singularity'' development on a vortex sheet in an inviscid fluid with no surface tension described by the Kelvin-Helmholtz instability, a higher-order system that shares linear growth rate dependence on $|k|$ \cite{Moore79,mbo82,Krasny86,Shelley92,cbt99}. We thus observe an identical initial growth behavior, but nonlinear terms for large amplitude waves result in a unique folding event in $t\approx 2.5$ in Fig.~\ref{fig:MainFig}b and very different long-time behavior.

\begin{figure}[th]
\begin{center}
\includegraphics[width=.48\textwidth]{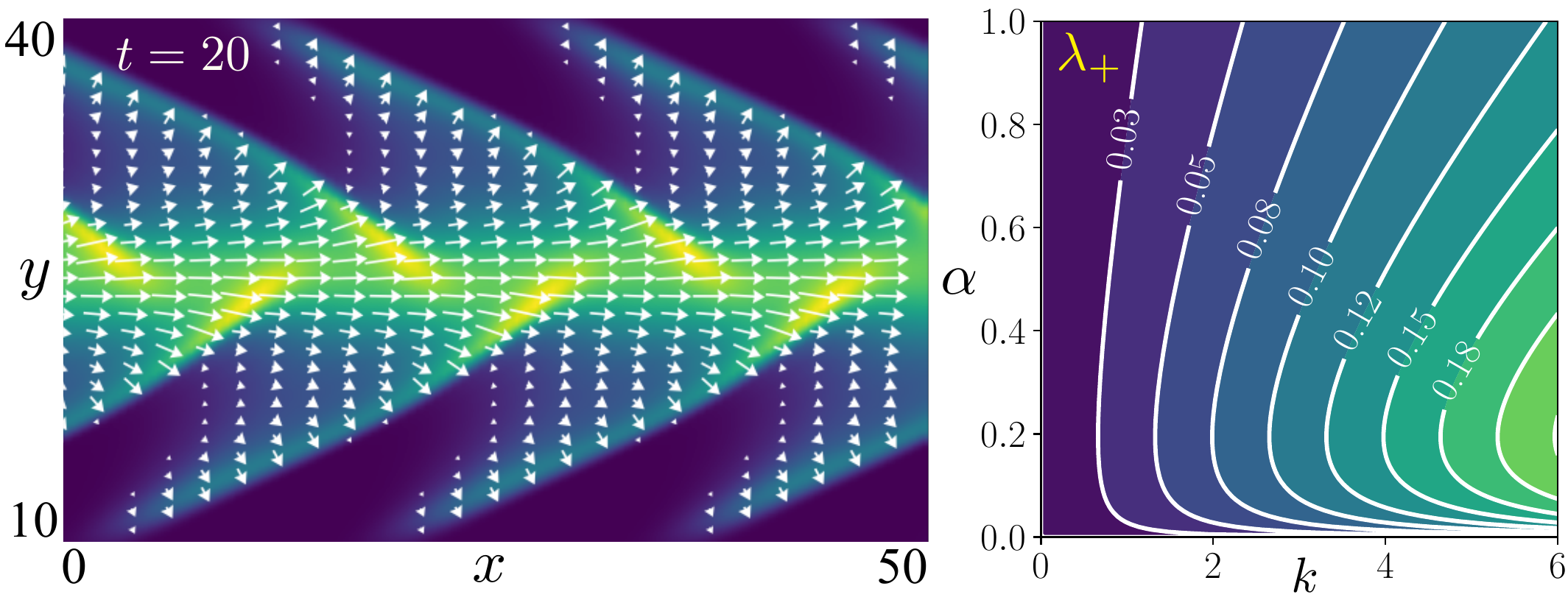}
\caption{(Left) Thin sheets of motile puller particles are unstable (here $\alpha=0.1$ and $d_t=d_r=0.001$). Arrows show the polarity field, $\langle \b{p} \rangle$. See Supplementary Movie M3. The initial condition is along an unstable eigenvector, $h(x,0)=0.144\cos(6\pi x/L)-0.063\sin(6\pi x/L))$ and $\phi(x,0)= 0.1\cos(6\pi x/L)$. (Right) Theoretical growth rates for motile pullers are non-monotonic in $\alpha$.}
\label{fig:Puller_instability}
\end{center}
\end{figure}
Meanwhile, in the motile case, $V_0=1$, sheets of pushers remain unstable for any dipole strength. Sheets of pullers, however, excite a positive-real-part eigenvalue in \eqref{eq:lambda}. Unlike in the case of pushers the maximal eigenvalue is not monotonic in the force dipole strength (Fig.~\ref{fig:Puller_instability}). Expanding about small $\alpha$, the largest eigenvalue is found when $\alpha = 2L/\kappa$, at which point $\mbox{Re}[\lambda_{+}]=\pi k(2L)^{-1}$ (since $\mbox{Re}[\lambda_{+}]\sim \pi k\alpha \kappa  (4 L^2)^{-1}$). For either motile or immotile pullers, the velocity field (oppositely signed to that illustrated in Fig.~\ref{fig:growth_rate}) rapidly damps the initial displacement, $h(x,t)$, and now rotates particles towards the direction {\it perpendicular} to the concentration band. Motility, however, allows the displacement and orientation fields to synchronize, leading ultimately to rapid growth of the concentration band amplitude and a large departure away from the initial profile, as shown in Fig.~\ref{fig:Puller_instability}.

%\begin{figure}[bh]
%\begin{center}
%\includegraphics[width=.48\textwidth]{Puller_Instability}
%\caption{Top: Thin sheets} of immotile puller particles are stable, but motile particles (shown here with $\alpha=0.1$ and $d_t=d_r=0.001$) are unstable. See Supplementary Movie M3. The initial condition is along the unstable eigenvector with wavenumber $k=3$, $\e\, h(x,0)=0.144\cos(6\pi x/L)-0.063\sin(6\pi x/L))$ and $\e\, \phi(x,0)= 0.1\cos(6\pi x/L)$. Growth rates are shown on the right. Bottom: Two circular isotropically oriented colonies expand radially and pass through each other without interacting. The active force $\b{f}_a=\nabla\cdot \bm{\Sigma}_a$ (right) is curl-free, modifying only the pressure and resulting in no fluid flow, $\b{u=0}$. This linear superposition generalizes to arbitrary initial concentration fields. See Supplementary Movie M4.}
%\label{fig:Puller_instability}
%\end{center}
%\end{figure}

For the motile suspensions above with non-zero dipole strength $\alpha$ there can be competing effects; in particular, if $\max(\mbox{Re}[\lambda_{\pm}]) < 0$ all solutions to the linear system eventually arrive at the stable base state, but if the system departs from the linearized region of phase space fast enough such solutions may not be realized in the fully nonlinear dynamics. This potential for departure is seen most clearly if the particles are not stress-generating: with $\alpha=0$ the wave amplitude grows linearly in time since any particles with nonzero initial orientation angle drift off without resistance along characteristic curves.

\noindent \textbf{Isotropic suspensions remain velocity free: a ``no-flow theorem'':} Assuming uniqueness of solutions to Eqs.~\eqref{eq:smoluchowski}-\eqref{eq:Stokes}, active suspensions of motile or immotile particles modeled by Eqs.~\eqref{eq:smoluchowski}-\eqref{eq:Stokes} which are initially isotropic in orientation, $\Psi(\b{x},\b{p},t=0)=\Psi_0(\b{x})$, result in no fluid flow, $\b{u}(\b{x},t)=\b{0}$, {\it everywhere and for all time}

\noindent {\it Sketch of the proof:} The proof assumes uniqueness of solutions for Eqs.~\eqref{eq:smoluchowski}-\eqref{eq:Stokes}, which was shown for two-dimensions by Chen \& Liu \cite{cl13}. Consider first the solution $\Psi^*$ to the Smoluchowski equation without velocity terms,
\begin{gather}
\Psi^*_t +V_0\mathbf{p} \cdot \nabla \Psi^* - d_t\nabla^2_\mathbf{x} \Psi^* - d_r\nabla^2_\mathbf{p} \Psi^*  = 0,
\end{gather}
with an initial condition which is isotropic in orientation. The velocity field generated by this solution, $\b{u}[\Psi^*]$, is given in Fourier space by 
\begin{gather}
\hat{\b{u}}_\b{k}[\Psi^*] = (8\pi |\b{k}|^2)^{-1}(\b{I} - \hat{\b{k}}\hat{\b{k}})\cdot \hat{\bm{\Sigma}}_a\cdot\b{k},\\
\hat{\bm{\Sigma}}_a\cdot\b{k} = \int_{D} \b{p}(\b{p}\cdot\b{k}) \hat{\Psi}_{\b{k}}^*(\b{p},t)\,d\b{p},
\end{gather}
where $\b{\hat{k}}=\b{k}/|\b{k}|$. Writing $\b{k}$ in a spherical (3D) or polar (2D) coordinate system about $\b{p}$ we find $\hat{\bm{\Sigma}}_a\cdot\b{k} = \lambda_{\b{k}}(t)\b{k}$ for some scalar function $\lambda_{\b{k}}(t)$. Hence $\hat{\b{u}}_\b{k}[\Psi^*] =\b{0}$ and then $\b{u}[\Psi^*] =\b{0}$. Since $\b{u}[\Psi^*] =\b{0}$, $\Psi^{*}$ also solves Eqs.~\eqref{eq:smoluchowski}-\eqref{eq:Stokes} with velocity terms included. By the uniqueness assumption we finally have that $\mathbf{u=0}$ everywhere and for all time for any initially isotropic distribution. A more detailed proof is included in the Supplementary Material. 

\begin{figure}[ht]
\begin{center}
\includegraphics[width=.48\textwidth]{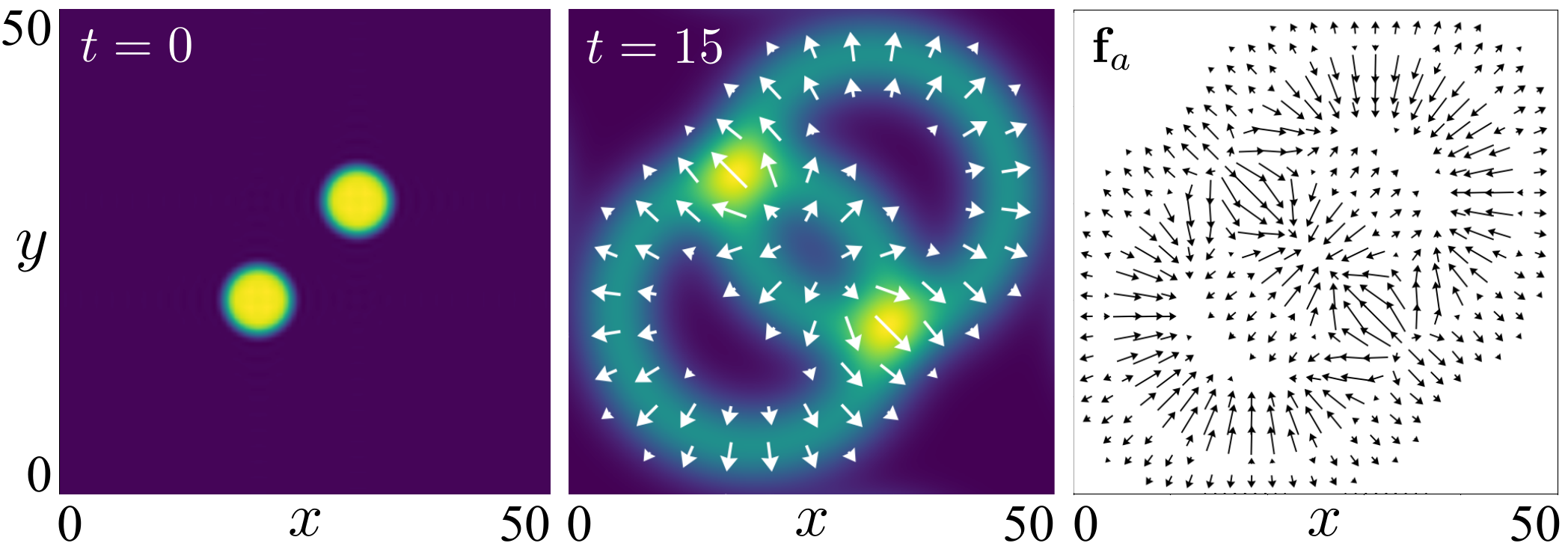}
\caption{The no-flow theorem playing out in a simple setting. Two circular isotropically oriented colonies expand radially and pass through each other without interacting. The polarity field $\langle \b{p} \rangle$ at $t=15$ shows local alignment. The active force $\b{f}_a=\nabla\cdot \bm{\Sigma}_a$ (right, at $t=15$) is curl-free, modifying only the pressure and resulting in no fluid flow, $\b{u=0}$ for all time. This linear superposition generalizes to arbitrary initial concentration fields. See Supplementary Movie M4.}
\label{fig:No_flow_theorem}
\end{center}
\end{figure}

The result is surprising since the system immediately loses orientational isotropy (see Fig.~\ref{fig:No_flow_theorem}), which would suggest the quick onset of a non-trivial flow field but this is not observed. Physically, the active force $\b{f}_a=\nabla \cdot \bm{\Sigma}_a$ is non-trivial for any $t>0$ but it is curl-free, so by the Helmholtz decomposition theorem $\b{f}_a = \nabla \lambda(t)$ for some scalar field $\lambda(t)$, which thus only modifies the pressure. As time progresses the force distribution evolves with the local active particle alignment, illustrated for two initially uniform colonies in Fig.~\ref{fig:No_flow_theorem}, but the expanding colonies simply pass through each other as linear waves. This behavior can be inferred even when including two-particle correlations \cite{snnmm17}.

Moreover, any distributions which result in $\b{u}=\b{0}$ for all time may be superimposed without generating a velocity field, for instance a random isotropic distribution may be perturbed by another distribution which has the property that $\nabla \cdot \langle \b{pp} \rangle = \nabla \chi(t)$ for any scalar $\chi(t)$, and still $\b{u=0}$ for all time. Physics which introduce nonlinearities in Eq.~\eqref{eq:smoluchowski}, such as near-field steric repulsion, are expected to nullify the theorem. 

%If instead of averaging to a mean-field theory we were to consider $N$ motile (or immotile) force dipoles with initially random orientation, the theorem states that as $N\to \infty$ the expected velocity at any point in space remains zero for all time, even after the particles have moved. This result can be inferred even when including two-particle correlations \cite{snnmm17}. 

The stability of the theorem to an initial localized alignment is not simply determined, as the initially isotropic state is not a stable base state. However, in light of the stability of the isotropic state of uniform concentration to large wavenumber perturbations~\cite{ss08} we expect an initial damping back towards isotropy. But on an extremely long timescale in a sufficiently large domain, the low wavenumber residue of the initial disruption is expected to lead to eventual growth along with a non-trivial flow. We have verified this prediction in at least one setting by numerical simulation, placing an aligned colony as in Fig.~\ref{fig:MainFig}a into a random concentration field which is orientationally isotropic. Persistent nematic alignment, for instance due to a boundary, may result in a more immediate transition to a global mean flow.

\noindent \textbf{Conclusion:} We have investigated colonies of active particles in the dilute regime as they invade a quiescent fluid. Colony-scale elongation depends on the sign of the active stress and can result in a self-buckling and self-folding cascade. Exponential growth at small times, linear in $|k|$, is mathematically equivalent to the Saffman-Taylor instability in a Hele-Shaw cell, or Rayleigh-Taylor instability in two-dimensional flow through a porous medium. The stability of sheets of pullers depends on particle motility with a growth rate which is non-monotonic in the dipole strength. Strikingly, a suspension modeled by pure far-field hydrodynamic interactions which is initially isotropic in orientation, even though isotropy is not preserved, results in no mean-field fluid flow everywhere and for all time. 

%However, a local perturbation to the initial isotropy results in self-straining, elongation, and generation of a global velocity field, just as shown for invasion of a quiescent fluid.

\noindent \textbf{Acknowledgements:} This project was initiated at the Woods Hole Oceanographic Institute as part of the Geophysical Fluid Dynamics summer program. Financial support is acknowledged by MJS (National Science Foundation Grants DMR-0820341 [NYU MRSEC], DMS-1463962, and DMS-1620331) and SES (DMR 767-1121288 [UW MRSEC] and DMS-1661900).

\bibliographystyle{apsrev}
\bibliography{BigBib}

\end{document}